\documentstyle[prl,aps,epsfig,color,amssymb]{revtex}

\begin{document}
\title{ 
       Spatio-temporal log-stable process for the \\
       turbulent energy-cascade
      }

\author{
J\"urgen Schmiegel$^{a+b}$
and Martin Greiner$^{a+c}$
}

\address{$^{a}$Max-Planck-Institut f\"ur Physik komplexer Systeme, 
               N\"othnitzer Str.\ 38, D--01187 Dresden, Germany}
\address{$^{b}$Department of Physics, University of Stellenbosch,
               7600 Stellenbosch, South Africa}
\address{$^{c}$Department of Physics, Duke University, Durham, NC 27708,
               USA}
\date{1.11.2001}

\maketitle

\vspace*{2cm}

\begin{abstract}
	We present a dynamical log-stable process for the spatio-temporal
        evolution of the energy-dissipation field in fully developed
	turbulence. The process is constructed from multifractal 
        scaling relations required for two-point correlators of arbitrary
        order. n-point correlation functions are calculated
        analytically and interpreted in terms of generalised fusion rules and
        in terms of the random multiplicative cascade picture. Multiplier
        distributions are compared with experimental results.
\end{abstract}

\vspace*{2cm}


\vspace*{4cm}

\noindent
CORRESPONDING AUTHOR: \\
J\"urgen Schmiegel \\
MPIPKS  \\
N\"othnitzer Str.\ 38 \\
Dresden 01187 \\ 
tel.:+49 (351) 871-1217 \\
fax: +49 (351) 871-1999  \\
email: schmiego@mpipks-dresden.mpg.de

\newpage
\section{Introduction}
The large number of strongly-coupled many-scales degrees of freedom in fully
developed turbulent flows renders any analytical and computational approach impractical for
very high Reynolds numbers. An alternative avenue is to look for a statistical description
\cite{YAG71} which by itself presents a difficult task. Only a few
rigorous results, as for example the $4/5$-law \cite{KOL41,FRI95}, have been
derived from the underlying Navier-Stokes equation, employing some idealised
assumptions like homogeneity, isotropy, etc. On the other hand, a multitude of
empirical facts are known from the experimental analysis
\cite{SRE97}, giving rise to a huge playground for phenomenological
modelling \cite{FRI95,BOHR98}.

One such class of data-driven models are random multiplicative cascade
processes (\emph{RMCP}) \cite{FRI95}, which are motivated by Richardsons energy cascade
picture
\cite{RICH22}. They emphasise
the observed multifractal nature of turbulent fields by assigning independent
and identically distributed random weights $q_{j}$ to a nested hierarchy of
inertial range scales $\eta \le l_{j} = L/ \lambda^{j} \le L$, confined by the
integral scale $L$ and the dissipation length scale $\eta$. Amplitudes $\Pi_{j}=q_{1}
\cdots q_{j}$ of the energy-flux density field at intermediate scales $l_{j}$
are then given as a multiplication of random weights, so that their respective
moments
\begin{equation}
\label{flux}
\left \langle \Pi^{n}_{j} \right \rangle = \left \langle q^{n}\right \rangle
^{j}= \left( \frac{L}{l_{j}}\right)^{\tau_{n}}
\end{equation}
indicate multiscaling. Note that interpretationally the energy flux field
$\Pi(x)$ is set equal to the energy-dissipation field $\epsilon(x)$ and that experimentally $\Pi_{j}$ is identified with
the coarse grained surrogate energy-dissipation field $\epsilon_{j}=l_{j}^{-1}
\int_{l_{j}} \epsilon(x)dx$. As a consequence of the scaling (\ref{flux}),
two-point correlations
\begin{equation}
\label{twopo}
\left \langle \epsilon(x_{1})^{n_{1}} \epsilon(x_{2})^{n_{2}}
\right \rangle\sim \left( \frac{L}{|x_{2}-x_{1}|}\right)^{\tau_{n_{1}+n_{2}}}
\left( \frac{|x_{2}-x_{1}|}{\eta}\right)^{\tau_{n_{1}}+\tau_{n_{2}}}
\end{equation}
also come with the same scaling \cite{YAG71}. It is not only the
multifractal aspect of these models which is confirmed by the data
\cite{MEN91}, more subtle observed quantities like multiplier
distributions \cite{SRE95,MOL95,PED96} and Markovian Kramers-Moyal
coefficients\cite{NAE} have also been
quantitatively reproduced by random multiplicative cascade processes
\cite{JOU99a,JOU00,JOU99b,CLE00}. But in spite of their success and inherent
simplicity, random
multiplicative cascade processes do have their drawbacks: at first there is no link to the
fundamental Navier-Stokes equation and, secondly, no dynamics is included in these
geometrical models as they are statically formulated in one-, sometimes in
multi-dimensional spaces only. In this work we do not address the first problem,
but present a solution for the second.

A dynamical generalisation of the geometric random
multiplicative cascade processes is proposed, which is continuous and causal
in space and time, and which generates positive-valued multifractal fields,
like the energy-dissipation field of fully developed turbulence. This paper is
an extension of Ref. \cite{let} (see also \cite{SCHM}). The
model construction is presented in Sect.\ II and relies on the multifractal scaling
form (\ref{twopo}) of two-point correlations only; stable statistics is used
to facilitate analytical results. $n$-point correlations are discussed in
Sect.\ III in the context of generalised fusion rules and are interpreted in
terms of the random
multiplicative weight picture of the geometric model precursors. Standard and
non-standard observables, like integral moments and multiplier distributions
are addressed in Sect.\ IV and shown to agree with
data. Some model variations are presented in Sect.\ V.
\section{model-construction based on two-point correlators}
\subsection{Ansatz}
For the positive-valued energy-dissipation field $\epsilon(x,t)$, depending on
one space and time coordinate $x$ and $t$, we assume a
spatio-temporal multiplicative process, 
\begin{eqnarray}
\label{def1}
\epsilon(x,t) & = & \exp \left \{ \int_{-\infty}^{\infty}\!dt'
\int_{-\infty}^{\infty}\!dx'f(x-x',t-t')\; \gamma(x',t') \right \}
\end{eqnarray} 
where $\gamma \sim S_{\alpha}((dx
dt)^{\alpha^{-1}-1} \sigma,-1,\mu)$ is distributed
according to a stable white-noise field \cite{TAQ94}. The normalisation
$\langle \epsilon(x,t) \rangle =\rm{const}=1$ determines the parameter
$\mu=\sigma^{\alpha}/\cos (\pi \alpha/2)$. The symmetric index function
\begin{eqnarray}
f(x,t)=\left \{ \begin{array}{ll} 1 & \;\;\;\;\;\;\; (0 \le t \le T, \; -g(T-t) \le x \le
g(T-t))  \\ & \\ 0 & \;\;\;\;\;\;\;  (\mbox{otherwise}) \end{array} \right.
\end{eqnarray}
is defined by the kernel-function $g(t)$; see Fig.1. Details of the explicit
functional form for $g(t)$ will be derived in Sect.\ II.B. 
The energy-dissipation field $\epsilon(x,t)$ at spatio-temporal position $(x,t)$ is composed multiplicatively of independent and identically
distributed (\emph{i.i.d.}) random-numbers $\exp
\left \{ \gamma(x,t) \right \}$, lying inside an attached domain,
bounded by the kernel-function $g(t)$. The condition $0 \le t \le T$ for the
index-function $f \neq 0$ ensures causality and defines the correlation-time
$T$ as 
the maximum of 
the temporal extension of the influence domain. $L$ gives the maximum of the
spatial extension and the correlation is given by $2g(0)=L$. With the
locality-condition $g(T)=0$, the influence domain is attached to the point
$(x,t)$ in an unequivocal way.
These boundary conditions for the kernel-function $g(t)$ make the
ansatz (\ref{def1}) a dynamical, causal and continuous process for the energy-dissipation
field. This process is also homogeneous in space and time, as we assumed an \emph{i.i.d}
random-field $\gamma(x,t)$ and independence of the form of the influence
domain from the 
spatio-temporal position.

The specification of the model is completed once the kernel-function $g(t)$
and the remaining parameters of the random-field $\gamma(x,t)$ are determined. This will be done in the
next subsection where multifractal statistics is demanded from the ansatz (\ref{def1}).
Technically two-point correlations will be employed for this derivation. Their
geometrical structure is illustrated in Fig.2 for various 
spatio-temporal distances. We see that the correlation structure is a function
of overlaps of influence domains, which on the other hand are bounded by the
kernel-function $g(t)$. Thus two-point correlations give direct information
about the kernel-function. Since we want to construct our process to produce  a multifractal
field, we first have to specify the interplay between two-point correlations
and multifractality.

Multifractality of the energy-dissipation field $\epsilon(x,t)$ in fully
developed turbulence \cite{FRI95} is usually
observed by examining the scaling behaviour $\left\langle \overline{\epsilon}_{l}(x,t)^{n} \right \rangle
\sim l^{-\tau_{n}}$ of the field amplitude  
$\overline{\epsilon}_{l}(x,t)=l^{-1}\int_{x-l/2}^{x+l/2} dx'\epsilon(x',t)$,
coarse grained over a
spatial domain of size $l$ with centre $x$ \cite{MEN91}; angular brackets denote the expectation value as the
average over independent realizations of the field. Multifractality prevails
once the scaling exponents $\tau_{n}$ are represented by polynomials in $n$,
although often for given $n$, it is already difficult to unambiguously extract a
unique value for $\tau_{n}$ via this standard route. Since the employed
moments involve an integration over
$n$-point statistics $\langle \epsilon(x_{1},t) \cdots
\epsilon(x_{n},t)\rangle$, the latter are more fundamental to reveal
multifractality. For example, an inertial range scaling $ \left \langle
\overline{\epsilon}_{l}^{2}(x,t) \right \rangle \sim l^{-\tau_{2}}$ of the
second-order integral moment automatically implies the same scaling $ \left
\langle \epsilon(x,t) \epsilon(x+l,t) \right \rangle \sim l^{-\tau_{2}}$ for
the two-point correlation; however the reverse need not be true
\cite{GRE96,WOL00}.

For our purpose, we use two-point spatio-temporal 
correlators of arbitrary positive orders
\begin{eqnarray}
\label{cort}
r_{n_{1},n_{2}}(\Delta x,\Delta t) & = & \frac {\left \langle
\epsilon^{n_{1}}(x,t)\epsilon^{n_{2}}(x+\Delta x,t+\Delta t)\right \rangle }{\left \langle
\epsilon^{n_{1}}(x,t)\right \rangle \left \langle \epsilon^{n_{2}}(x+\Delta x,
t+\Delta t)\right \rangle}. 
\end{eqnarray}
Spatio-temporal
multifractality is incorporated by  demanding scaling relations for the temporal and
spatial two-point correlators 
\begin{eqnarray}
\label{corrt}
r_{n_{1},n_{2}}(\Delta x=0,\Delta t) & \sim  &  \Delta t^{-\tau[n_{1},n_{2}]} \;\;\;\;\;\;\;
(t_{L} \le \Delta t \le T-t_{\eta})\\ \nonumber \\
\label{corrs}
r_{n_{1},n_{2}}(\Delta x, \Delta t=0) & \sim  &  \Delta x^{-\overline{\tau}[n_{1},n_{2}]} \;\;\;\;\;\;\;
(\eta \le \Delta x \le x_{L})
\end{eqnarray}
with temporal and spatial multifractal scaling exponents $\tau[n_{1},n_{2}]$
and $\overline{\tau}[n_{1},n_{2}]$. For future purposes, we introduce arbitrary
temporal $\Delta t \in [t_{L},T-t_{\eta}]$  
and spatial boundaries $\Delta x \in [\eta,x_{L}]$ for the scaling relations
(\ref{corrt}) and (\ref{corrs}) to hold. One can
think of $[\eta,x_{L}]$ as defining the inertial range for two-point
correlators. Since we are dealing with a dynamical model,
$[t_{L},T-t_{\eta}]$ then defines a temporal inertial range.
\subsection{Determination of the kernel-function ${\bf{\boldmath g(t)}}$}
The multiplicativity of the process (\ref{def1}) with a stable white-noise
field and the multifractal scaling
of two-point correlators (\ref{corrt}) and (\ref{corrs}) are the only
ingredients of our ansatz. The functional form of the kernel-function $g(t)$ 
will now be deduced solely out of these premises. Fig.2a illustrates the
overlap of the influence domains of two points of the energy-dissipation field, having the same
spatial 
position $x$ but separated by a temporal distance $\Delta t$ with $t_{L} <
\Delta t <
T-t_{\eta}$. Using the independence of $\gamma(x,t)$ at
different spatio-temporal positions, the temporal two-point
correlator (\ref{corrt}) becomes
\begin{eqnarray}
r_{n_{1},n_{2}}(\Delta x=0,\Delta t) & = & 
\frac{ \left \langle \exp \left \{ (n_{1}+n_{2})\int_{V(\Delta t)} dx'
dt' \gamma(x',t') \right \} \right \rangle}{\left \langle \exp \left \{
n_{1}\int_{V(\Delta t)} dx'
dt' \gamma(x',t') \right \} \right \rangle \left \langle \exp \left \{
n_{2}\int_{V(\Delta t)} dx'
dt' \gamma(x',t') \right \} \right \rangle}.
\end{eqnarray}
The influence domains of both points of the energy-dissipation field factorise
into an overlap contribution $V(\Delta t)$ and two non-overlap
contributions $V(0)-V(\Delta t)$. The latter appears in the denominator and
the numerator of
(\ref{cort},) in the same 
respective orders $n_{1}$ and $n_{2}$ and cancel out. The remaining contributions are
those from the overlap of the two influence domains. The overlap $V(\Delta t)$ is bounded by the
kernel-function $g(t)$.

Using the Laplace-transform and the
stable-property
\begin{equation}
\left \langle \exp \left \{\lambda_{1} \gamma_{1}+\lambda_{2}
\gamma_{2} \right \} \right \rangle=\exp
\left \{\frac{-\sigma^{\alpha}}{\cos \left( \frac{\pi \alpha}{2}\right)} \left
(\lambda_{1}^{\alpha}+\lambda_{2}^{\alpha} \right )+\mu \left(
\lambda_{1}+\lambda_{2} \right )\right \}
\end{equation}
with \emph{i.i.d.} $\gamma_{1},\gamma_{2}\sim S_{\alpha}(\sigma,-1,\mu)$
and $\lambda_{1},\lambda_{2}\ge 0$,
together with $\lambda
\gamma \sim S_{\alpha}(|\lambda| \sigma,\beta,0)$ and $\gamma+\mu \sim
S_{\alpha}(\sigma,\beta,\mu)$ for $\gamma \sim S_{\alpha}(\sigma,\beta,0)$,
after a straightforward calculation we get for the temporal correlators 
\begin{eqnarray}
\label{time}
r_{n_{1},n_{2}}(\Delta x=0,\Delta t) & = & \exp \left \{ \frac{-\sigma^{\alpha}}{\cos \left (
\frac{\pi \alpha}{2}\right)} \left (
\left(n_{1}+n_{2}\right)^{\alpha}-n_{1}^{\alpha}-n_{2}^{\alpha}\right)
V(\Delta t)
\right \}.
\end{eqnarray}
Eq.\ (\ref{time}) is a purely geometrical relation connecting two-point
statistics with overlapping domains. Since these overlap-domains are bounded by
the kernel-function $g(t)$, the scaling relation (\ref{corrt}) translates into
\begin{eqnarray}
\label{vol}
V(\Delta t) & = & \int_{\Delta t}^{T} dt' \int_{-g(t')}^{g(t')} dx'\sim \ln
\Delta t
\end{eqnarray}
for $t_{L} \le \Delta t \le T-t_{\eta} $. Combining (\ref{vol})
with (\ref{corrt}) and 
(\ref{time}), the temporal scaling exponent then becomes
$\tau[n_{1},n_{2}]=a((n_{1}+n_{2})^{\alpha}-n_{1}^{\alpha}-n_{2}^{\alpha})$,
with some constant $a$.  
Finally after differentiating (\ref{vol}) with respect to $\Delta t$, we get the kernel-function
\begin{eqnarray}
\label{win1}
g(t) & = & \frac{a_{0}}{t}
\end{eqnarray}
inside the temporal scaling regime $t \in [t_{L},T-t_{\eta}]$, where $a_{0}=\frac{a \cos
\left( \frac{\pi \alpha}{2}\right)}{-2\sigma^{\alpha}}$.
The singularity of $g(t)$ at $t\rightarrow 0$ and the condition of locality $g(T)=0$ now
justifies the introduction of the boundaries $t_{L}$ and
$T-t_{\eta}$ for the temporal scaling regime.

The scaling relation (\ref{corrt}) is valid independent of the
behaviour of $g(t)$ for $t>T-t_{\eta}$. This part of $V(\Delta t)$ in Fig. 2a
is always a complete part of the overlap as long as $t_{L} \le \Delta t \le
T-t_{\eta}$ and therefore contributes only as a
constant prefactor. This gives the possibility to achieve locality $g(T)=0$
with the arbitrary choice
\begin{eqnarray}
\label{win2}
g(t) & = & \frac{a_{0}}{t+\tilde{t}}-\frac{a_{0}}{T+\tilde{t}} \;\;\;\;\; \;\;\;\;\;
\left(T-t_{\eta} \le t \le T\right),
\end{eqnarray}
where  at $t=T-t_{\eta}$, $\tilde{t}=\sqrt{T t_{\eta}-7t_{\eta}^{2}/4}-T+t_{\eta}/2$ is
fixed by continuity with (\ref{win1}). For $\tilde{t}$ real, we have to
guarantee $T>7t_{\eta}/4$.    
The reason for the specific choice (\ref{win2}) is
simply to stay in the same class of polynomials of order $-1$ and to be able
to give an analytical expression for the new constant $\tilde{t}$. 

The temporal scaling behaviour is independent of the functional form of $g(t)$ for $0 \le t \le t_{L}$, as the intersection domain $V(\Delta t)$ is not
bounded by $g(t)$ for $t<t_{L}$. This domain is fixed by the equal time
two-point correlators 
\begin{eqnarray}
\label{correl}
r_{n_{1},n_{2}}(\Delta x,\Delta t=0) & = & \frac{ \left \langle \exp \left \{
(n_{1}+n_{2})\int_{V(\Delta x)} dx'
dt' \gamma(x',t') \right \} \right \rangle}{\left \langle \exp \left \{
n_{1}\int_{V(\Delta x)} dx'
dt' \gamma(x',t') \right \} \right \rangle \left \langle \exp \left \{
n_{2}\int_{V(\Delta x)} dx'
dt' \gamma(x',t') \right \} \right \rangle},
\end{eqnarray}
characterised by a spatial intersection domain
$V(\Delta x)$
and calculated in the same way as for temporal correlators; consult again
Fig. 2b.
For simplicity the boundaries $\eta=2g(T-t_{\eta})$ and
$x_{L}=2g(t_{L})$ of the scaling regime (\ref{corrs}) are set equal to those
of the temporal two-point correlators. Then the relation
\begin{equation}
\label{timei}
2g(t(\Delta x))=\Delta x
\end{equation}
yields the intersection time $t(\Delta x)$,
where the boundaries of the two domains of influence intersect with each other. Making use of (\ref{win1}), the spatial
overlap reads 
\begin{equation}
V(\Delta x)=\int_{0}^{t(\Delta x)}dt' (2g(t')-\Delta x)=
2\int_{0}^{t_{L}}dt'g(t')+2a_{0}\left( \ln \frac{2a_{0}}{\Delta xt_{L}}-1\right). 
\end{equation}
The simple choice 
\begin{eqnarray}
g(t) & = & \frac{2a_{0}}{t+t_{L}}\;\;\;\;\; \;\;\; \left(0 \le t \le t_{L}\right)
\end{eqnarray}
is not unique, but is acceptable as a simple continuous extension of
(\ref{win1}) into the domain $0 \le t \le
t_{L}$, leading also to the perfect scaling relation (\ref{corrs}). Finally the
cut-off $t_{L}$ is fixed by the condition $g(0)=L/2$. The kernel-function
$g(t)$ is completely specified by the scaling relations (\ref{corrt}) and (\ref{corrs}). It
comes with the four parameters $T$, $L$, $t_{L}$, $t_{\eta}$, with the first two
representing the correlation time and length and the latter two limiting the
perfect scaling range of the two-point correlators. Note that by construction, the limit of the temporal scaling regime $t_{\eta} \neq 0$ is necessary for
locality $g(T)=0$ and avoids ultraviolet divergence in the spatial
correlations. Two more parameters
are hidden in the field construction (\ref{def1}) and enter via the stable
white-noise field; these are the stable index $\alpha$ and the parameter $a$,
determining the absolute strength of the multifractal scaling exponents. By
adjusting to the experimentally observed scaling exponents
$\tau[1,1]=\tau_{2}=0.225$ \cite{SRE97}, one of them, say
$a=0.225/(2^{\alpha}-2)$ becomes expressible in terms of the other.
Fig.\ 1 
shows the kernel-function $g(t)$ for parameters $\alpha=1.9$ and $T=L=2^{8}$,
$t_{L}=t_{\eta}=10$ in arbitrary units. Note again the three
different functional forms for the kernel-function $g(t)$, the arbitrary small-scale contribution for $t>T-t_{\eta}$, the contribution $t_{L} \le t \le
T-t_{\eta}$ responsible for scaling of temporal two-point correlators and the
large scale contributions $t \le t_{L}$ responsible for the perfect scaling of spatial
two-point correlators.

Independent of a special choice of parameters, the presented solution for $g(t)$
implies spatio-temporal equivalence of the scaling exponents
\begin{equation}
\label{exp}
\overline{\tau}[n_{1},n_{2}]=\tau[n_{1},n_{2}]=a\left( \left(n_{1}+n_{2}\right)^{\alpha}-n_{1}^{\alpha}-n_{2}^{\alpha}\right). 
\end{equation}
This corresponds to the
Taylor-Frozen-Flow Hypothesis \cite{YAG71} where correlations at a spatial distance are
measured at a temporal distance. Here this hypothesis holds at least in the case of two-point
correlations inside the scaling regime. Later on, it turns out that this is not
true for arbitrary spatio-temporal $n$-point correlations.  
\subsection{Link to RMCP}
In Ref.\ \cite{let} the kernel-function $g(t)$ has been constructed in a different
approach by linking the dynamical
ansatz (\ref{def1}) directly to geometric random multiplicative cascade
processes (RMCP). The functional form of the kernel-function $g(t)$ in \cite{let}
is more or less the same, differing only in the regimes $0 \le t \le t_{L}$ and
$T-t_{\eta} \le t \le T$.

In a discrete RMCP the energy-dissipation 
\begin{eqnarray}
\label{rmcm}
\epsilon(\eta)=\prod_{j=1}^{J}q(l_{j})=\exp \left \{\sum_{j=1}^{J} \ln
q(l_{j})\right \} 
\end{eqnarray}
resolved at the
dissipation scale $\eta$ is given as the product of \emph{i.i.d.} multiplicative
weights $q(l_{j})$ of a nested hierarchy of scales $\eta \le l_{j}=L'/\lambda^j
\le L'$ with $\lambda>1$.
Comparing (\ref{rmcm}) with (\ref{def1}), the scale index $j$ is replaced by a
discrete time-index $t_{j}$ with $2g(t_{j})=l_{j}$, leading to
\begin{equation}
\label{weight}
q(l_{j})=\exp \left \{ \int_{t_{j-1}}^{t_{j}} dt' \int_{-g(t')}^{g(t')} dx'
\gamma(x',t') \right \}.
\end{equation}
Since $\gamma(x,t)$ is a stable white noise field, the independent
multiplicative weights $q(l_{j})$ become identically distributed once the
volume associated to the time-strip integration, inside the exponential is
scale-, i.e. $j$-independent. This leads directly to the result (\ref{win1}), valid for $t_{L}=g^{-1}(L'/2) \le t \le T-t_{\eta}$.

Identifying
\begin{eqnarray}
\label{weight1}
\tilde{q}(L) & = & \exp \left \{ \int_{0}^{t_{L}} dt' \int_{-g(t')}^{g(t')} dt'
\gamma(x',t')\right \}
\end{eqnarray}
and
\begin{eqnarray}
\label{weight2}
\tilde{q}(\eta) & = & \exp \left \{ \int_{T-t_{\eta}}^{T} dt' \int_{-g(t')}^{g(t')} dt'
\gamma(x',t')\right \} 
\end{eqnarray}
as large-scale and small-scale fluctuations, respectively, the field
(\ref{def1}) can be written in the RMCP-analogue form
\begin{equation}
\epsilon(x,t)=\tilde{q}(\eta) \left(\prod_{j=1}^{J}q(l_{j})\right) \tilde{q}(L),
\end{equation}
where $J=\log_{\lambda}\left( \left(T-t_{\eta}\right)/t_{L}\right)$
represents the number of discrete cascade steps.

Note that all $q(l_{j})$ are \emph{i.i.d.} for $1 \le j \le J$, whereas $\tilde{q}(\eta)$
and $\tilde{q}(L)$ come with different distributions. There is more analogy to
geometric RMCP: in Ref.\cite{SCH}, it has been demonstrated that the
experimentally motivated unrestrictive sampling of spatial two-point
correlators leads to finite-size corrections to scaling, once the two-point
distance approaches the integral length $L$ and that this finite-size
correction can be removed by a suitably tuned large-scale fluctuation. Since
by construction, the dynamical RMCP (\ref{def1}) comes with perfect scaling for
the two-point correlators (\ref{corrs}), the large scale fluctuation $\tilde{q}(L)$
has to be viewed as perfectly tuned.
\section{More on n-point correlations}
\subsection{Spatio-temporal two-point correlations}
In the previous section we used only spatial and temporal  correlations to fix the
kernel-function $g(t)$.  Now we examine arbitrary spatio-temporal two-point 
correlations $\langle \epsilon(x,t) \epsilon(x+\Delta x,t+\Delta t) \rangle$.  
It is straightforward to calculate these correlations as a function of the
spatio-temporal overlap $V(\Delta x,\Delta t)$, which is shown as the shaded
area in Fig. 2c. Analytical results however
need a proper distinction of several cases, which depend on the relative position of the two
space-time points. All can be treated
analytically. Instead of presenting all these formulas, we prefer
only to illustrate them with Fig.3.
Fig.3a shows the two-point correlations as a function
of $\Delta t$ and $\Delta x$. Parameters are $\alpha=1.9$, $\tau[1,1]=0.225$
and in arbitrary units 
$t_{L}=5$, $t_{\eta}=1$, $T=L=2^{8}$. For $\Delta x=0$ a rigorous
straight-line behaviour is observed in this log-log-log plot for $t_{L} \le \Delta t \le T-t_{\eta}$,
which reflects the perfect scaling form of the limiting case
(\ref{corrt}). The same holds for the limiting case $\Delta t=0$ in the range
$\eta \le \Delta x \le x_{L}$; see again Eq. (\ref{corrs}). As $\Delta x$
and/or $\Delta t$ increase, more and more deviations from the rigorous scaling
behaviour can be seen. Once $\Delta x > L/2 +g(\Delta t)=L_{corr}(\Delta t)$,
the two influence domains do not overlap any more and as a consequence
decorrelation $\left \langle \epsilon(x,t)\epsilon(x+\Delta x \ge x+L_{corr}(\Delta t),t+\Delta t)\right \rangle=\left \langle
\epsilon(x,t)\right \rangle \left \langle \epsilon(x+\Delta x,t+\Delta
t)\right \rangle$ sets in. It is interesting to note that $L_{corr}(\Delta
t)$ displays the functional form of the kernel-function $g(t)$.
\subsection{Equal-time n-point correlations I}
On our way to more complex correlation functions, equal time
$n$-point correlations 
\begin{eqnarray}
\rho_{n}(x_{1},m_{1}; \ldots ;x_{n},m_{n})& = & \left \langle
\epsilon(x_{1},t)^{m_{1}} \cdots 
\epsilon(x_{n},t)^{m_{n}} \right \rangle
\end{eqnarray}
of arbitrary order $m_{1}, \ldots ,m_{n}$ are now investigated.
First, we will reveal their general structure. In the two following subsections they will be discussed in the context of fusion rules \cite{PRO96}
and generic RMCP structures. We restrict ourselves to equal-time correlations. The case
of temporal correlations is straightforward and does not give any new
insight. The more complicated inspection of arbitrary spatio-temporal $n$-point correlation
functions is briefly touched  in Subsect. III.E. 

We again start with two points to exhibit the transition from two-point
correlators to two-point correlation densities. Making use of (\ref{cort})
and the result (\ref{correl}), the equal-time two-point correlation density
becomes: 
\begin{eqnarray}
\label{cd2}
\rho(x_{1},m_{1};x_{2},m_{2}) & = & \left \langle \epsilon(x_{1},t)^{m_{1}}
\epsilon(x_{2},t)^{m_{2}} \right \rangle \nonumber \\ & & \nonumber \\
 & = & r_{m_{1},m_{2}}(|x_{2}-x_{1}|,\Delta
t=0) \left \langle \epsilon(x_{1},t)^{m_{1}}\right \rangle \left \langle
\epsilon(x_{2},t)^{m_{2}}\right \rangle \nonumber \\ & & \nonumber \\ & = & \exp
\left \{ \frac{\tau_{2}}{2a_{0}\left(2^{\alpha}-2\right)} \left(\left (
m_{1}^{\alpha}-m_{1}\right)\left(V_{1}/ V_{2}\right)+\left(
m_{2}^{\alpha}-m_{2}\right)\left(V_{2}/ V_{1}\right) \right.\right. \nonumber
\\ & & \nonumber \\
& &
\left.
\left.\;\;\;\;\;\;\;\;+\left(\left(m_{1}+m_{2}\right)^{\alpha}-m_{1}-m_{2}\right)\left
(V_{1}\cap
V_{2}\right)
\right)\right\} \nonumber 
\\ & &  \nonumber \\
 & = & \overline{\rho}(m_{1}|V_{1}/ V_{2})\overline{\rho}(m_{2}|V_{2}/ V_{1})
\overline{\rho}(m_{1}+m_{2}|V_{1} \cap V_{2}).
\end{eqnarray}
Here $V_{i}=\int_{0}^{T}dt \int_{-g(t)}^{g(t)} dx$ represents the volume of
the influence domain, attached to point $i$. $V_{i} / V_{j}$
represents volume $V_{i}$ without the overlapping volume $V_{i} \cap V_{j}$
resulting from $V_{j}$. In Fig. 4a the volumes $V_{1} / V_{2}$, $V_{2}
/ V_{1}$ and $V_{1} \cap V_{2}$ are labeled with their respective
orders $m_{1}$, $m_{2}$ and $m_{1}+m_{2}$; these three disjunct contributions
fill the union of the  two influence domains with no remaining overlaps and consequently
factorise, leading to the last step of (\ref{cd2}), where the abbreviation 
\begin{equation}
\label{abbre}
\overline{\rho}(m|V)=\exp \left \{ \frac{\tau_{2}}{2a_{0}\left(2^{\alpha}-2\right)} \left (
m^{\alpha}-m \right) V \right \}
\end{equation}
has been introduced.

For the third order $n=3$ with points $x_{1} < x_{2} <x_{3}$, we arrive at the
straightforward generalisation of (\ref{cd2})
\begin{eqnarray}
\label{cd3}
\rho(x_{1},m_{1};x_{2},m_{2};x_{3},m_{3}) & = & 
\overline{\rho}(m_{1}|V_{1} / V_{2})\; \overline{\rho}(m_{2}|V_{2}
/ (V_{1} \cup V_{3}))\; \overline{\rho}(m_{3}|V_{3} / V_{2})
\nonumber \\ & & \nonumber \\ & & \overline{\rho}(m_{1}+m_{2}|(V_{1} \cap
V_{2}) / V_{3})\; \overline{\rho}(m_{2}+m_{3}|(V_{2} \cap
V_{3}) / V_{1})\nonumber \\ & & \nonumber \\ & & \overline{\rho}(m_{1}+m_{2}+m_{3}|V_{1} \cap V_{2}
\cap V_{3});
\end{eqnarray}    
Consult Fig. 4b. For the most general case of an equal-time $n$-point
correlation density with $x_{1} < x_{2} < \ldots < x_{n}$ we only state the result:
\begin{eqnarray}
\label{sum}
\rho(x_{1},m_{1}; \ldots ;x_{n},m_{n}) & = &\prod_{j=1}^{n}
\prod_{k=1}^{n-j+1}\overline{\rho}(m_{k}+ \ldots +m_{k+j-1}|(V_{k} \cap \ldots
\cap V_{k+j-1}) / (V_{k-1} \cup V_{k+j})).
\end{eqnarray}
One gets contributions of all common histories of neighbouring
points, first the contributions every point has non-overlapping with the
others, second all contributions from overlaps with one neighbour, with two
neighbours,$\ldots$  and finally one contribution of the overlap, all points
have in common. 
\subsection{Equal-time n-point correlations II: generalised fusion rules}
After presenting the overall structure of equal-time $n$-point
correlation densities, in terms of the intuitive picture with disjunct
contributions from different overlap volumes of the respective influence
cones, we now aim to rewrite them in terms of a fusion-rule picture. Again,
the focus is first on two-point, then on three- and four-point, and finally on
$n$-point correlation densities.

Since the one-point moment $\left \langle \epsilon (x,t)^{m} \right \rangle$
is independent of the position $x$, the scaling behaviour of the equal-time
two-point correlation density $\rho(x_{1},m_{1};x_{2},m_{2})$ is identical to
the two-point correlator (\ref{corrs}) for $\eta \le \Delta x
=x_{2}-x_{1} \le x_{L}$. Hence we arrive at 
\begin{eqnarray}
\label{fusion1}
\rho(x_{1},m_{1};x_{2},m_{2}) & \sim & \left (
x_{2}-x_{1}\right)^{-\xi[m_{1},m_{2}]}  
\end{eqnarray}
with 
\begin{equation}
\label{fusion1.1}
\xi[m_{1},m_{2}]=\frac{\tau_{2}}{2^{\alpha}-2} \left(\left (m_{1}+m_{2}\right)^{\alpha}-m_{1}^{\alpha}-m_{2}^{\alpha}\right).
\end{equation}
For our purposes it is instructive to rederive this result from the last line
of (\ref{cd2}). Noting that $\overline{\rho}(m|V_{1} /
V_{2})=\overline{\rho}(m|V_{1})/\overline{\rho}(m|V_{1} \cap V_{2})$ the
scaling part of (\ref{cd2}) is given by
\begin{eqnarray}
\label{fusion2}
\rho(x_{1},m_{1};x_{2},m_{2}) & \sim & \frac{\overline{\rho}(m_{1}+m_{2}|V_{1}
\cap V_{2})}{\overline{\rho}(m_{1}|V_{1} \cap V_{2}) \overline{\rho}(m_{2}|
V_{1} \cap V_{2})}.
\end{eqnarray}
Realizing that $V_{1} \cap V_{2}= V(\Delta x =x_{2}-x_{1}) \sim \ln \Delta x$
and making use of (\ref{abbre}), Eq.\ (\ref{fusion2}) then transforms into
Eq.\ (\ref{fusion1}). 

In the expression (\ref{cd3}) for the equal-time
three-point correlator each of the six contributing factors is rewritten and
then reordered, giving rise to 
\begin{eqnarray}
\label{fusion3}
\lefteqn{\rho(x_{1},m_{1};x_{2},m_{2};x_{3},m_{3})=} \nonumber \\  \nonumber \\
&=&\frac{\overline{\rho}(m_{1}|V_{1})}{\overline{\rho}(m_{1}|V_{1} \cap V_{2})}
\; \frac{\overline{\rho}(m_{2}|V_{2}) \overline{\rho}(m_{2}|V_{1} \cap
V_{3})}{\overline{\rho}(m_{2}|V_{1} \cap V_{2}) \overline{\rho}(m_{2}|V_{2}
\cap V_{3})}\; \frac{\overline{\rho}(m_{3}|V_{3})}{\overline{\rho}(m_{3}|V_{2} \cap
V_{3})} \nonumber \\  \nonumber \\
&=& \frac{\overline{\rho}(m_{1}+m_{2}|V_{1} \cap
V_{2})}{\overline{\rho}(m_{1}+m_{2}|V_{1}\cap V_{3})}\; \frac{\overline{\rho}(m_{2}+m_{3}|V_{2} \cap
V_{3})}{\overline{\rho}(m_{2}+m_{3}|V_{1} \cap V_{3})}\;
\overline{\rho}(m_{1}+m_{2}+m_{3}|V_{1} \cap V_{3}) \nonumber \\ \nonumber
\\ &\sim& \frac{\overline{\rho}(m_{1}+m_{2}|V_{1} \cap
V_{2})}{\overline{\rho}(m_{1}|V_{1} \cap V_{2}) \overline{\rho}(m_{2}|V_{1}
\cap V_{2})}\; \frac{\overline{\rho}(m_{2}+m_{3}|V_{2} \cap
V_{3})}{\overline{\rho}(m_{2}|V_{2} \cap V_{3}) \overline{\rho}(m_{3}|V_{2}
\cap V_{3})} \nonumber \; \frac{\overline{\rho}(m_{1}+m_{2}+m_{3}|V_{1} \cap V_{3})
\overline{\rho}(m_{2}|V_{1} \cap V_{3})}{\overline{\rho}(m_{1}+m_{2}|V_{1}
\cap V_{3}) \overline{\rho}( m_{2}+m_{3}|V_{1} \cap V_{3})} \nonumber \\ 
\nonumber \\
&\sim& \left( x_{2}-x_{1}\right)^{-\xi[m_{1},m_{2}]} \left(
x_{3}-x_{2}\right)^{-\xi[m_{2},m_{3}]} \left(x_{3}-x_{1} \right)^{-\tilde{\xi}[m_{1},m_{2},m_{3}]}.
\end{eqnarray}
The exponents $\xi[m_{1},m_{2}]$ and $\xi[m_{2},m_{3}]$ are the same as in
(\ref{fusion1.1}), whereas the third one is now given by
$\tilde{\xi}[m_{1},m_{2},m_{3}]=\frac{\tau_{2}}{2^{\alpha}-2}\left((m_{1}+m_{2}+m_{3})^{\alpha}-(m_{1}+m_{2})^{\alpha}-(m_{2}+m_{3})^{\alpha}+m_{2}^{\alpha}\right)$.
The last line of (\ref{fusion3}) can be viewed as a generalised fusion
rule; consult also Ref.\cite{PRO96}.

The fusion rule (\ref{fusion3}) is easily generalised to arbitrary orders
$n$. This is possible because all intersection
domains can be written as a combination of $V_{i} \cap V_{j} \sim \ln
|x_{j}-x_{i}|$, given that $\eta < |x_{j}-x_{i}|<x_{L}$ for all pairs of
points. Proven by complete induction the equal-time $n$-point correlation
density has the following structure:
\begin{eqnarray}
\label{fusg}
\rho(x_{1},m_{1}; \ldots ;x_{n},m_{n}) & \sim & \left( \prod_{i=1}^{n-1}
\left(x_{i+1}-x_{i}\right)^{-\xi[m_{i},m_{i+1}]} \right) \prod_{j=2}^{n-1}
\prod_{l=j+1}^{n} \left(x_{l}-x_{l-j}\right)^{-\tilde{\xi}[m_{l-j}, \ldots ,
m_{l}]},
\end{eqnarray}
where
\begin{eqnarray}
\tilde{\xi}[m_{1}, \ldots, m_{i}] & = & \xi[m_{1}+ \ldots
+m_{i-1},m_{i}]-\xi[m_{2}+ \ldots +m_{i-1},m_{i}]\nonumber \\ & &\nonumber \\
 & = & \frac{\tau_{2}}{2^{\alpha}-2}\left(\left(m_{1}+ \ldots +m_{i}\right)^{\alpha}-\left( m_{1}+\ldots
+m_{i-1}\right)^{\alpha}-\left(m_{2}+\ldots
+m_{i}\right)^{\alpha}+\left(m_{2}+\ldots +m_{i-1}\right)^{\alpha}\right)
\end{eqnarray}
and $x_{1}<x_{2}< \ldots <x_{n}$ with $\eta <x_{i+1}-x_{i}<x_{L}$.
The equal-time $n$-point correlation function factorises into contributions at the
smallest scales $x_{i+1}-x_{i}$, next smaller scales $x_{i+2}-x_{i}$ $\ldots$
and finally one contribution at the largest scale $x_{n}-x_{1}$. The first
product in (\ref{fusg}) gives the contributions from the smallest scales
involving the scaling-exponents $\xi$ and the second product counts all
other scales with modified scaling exponents
$\tilde{\xi}$. These modified scaling-exponents arise from the nested
structure of the overlapping volumes.
\subsection{Equal-time n-point correlations III: RMCP interpretation}
So far the generalised fusion rule of the equal-time $n$-point correlation
densities has been expressed in two ways: Eq.\ (\ref{fusg}) uses relative
distances between the involved points, whereas Eq.\ (\ref{sum}) employs
overlap volumes of respective influence domains. Since already in Section\
II.C we interpreted the influence domains in terms of random
multiplicative weights, it is natural to interpret the generalised fusion rules
also in terms of random multiplicative weights. In order to simplify the
presentation, small- and large-scale fluctuations (\ref{weight1}) and
(\ref{weight2}) will be discarded.

Using the assignment (\ref{weight}) the expression (\ref{cd2}) for the
two-point correlation density can be expressed as
\begin{eqnarray}
\label{mul1}
\rho(x_{1},m_{1};x_{2},m_{2}) & = & 
\left \langle q^{m_{1}} \right \rangle^{j_{1}}
\left \langle q^{m_{2}} \right \rangle^{j_{2}}
\left \langle q^{m_{1}+m_{2}} \right \rangle^{j_{1,2}}.
\end{eqnarray}
In the RMCP language,
\begin{equation}
\label{j12}
j_{1,2}=\frac{V_{1} \cap V_{2}}{V_{\lambda}}
\end{equation}
with
\begin{equation}
V_{\lambda}=\int_{t_{j-1}}^{t_{j}}dt' \int_{-g(t')}^{g(t')}dx' = 2a_{0} \ln \lambda.
\end{equation}
denoting the number of steps the two points have evolved together during the
cascade history and
\begin{eqnarray}
\label{j1j2}
j_{1}=j_{2}=\frac{V_{1} / \left(V_{1} \cap V_{2}
\right)}{V_{\lambda}}= \frac{V_{2} / \left(V_{1} \cap V_{2}
\right)}{V_{\lambda}}  
\end{eqnarray} 
represent the number of cascade steps the two points go through independently
after their branching.

The three-point correlation density (\ref{cd3}) displays the same structure:
\begin{eqnarray}
\label{three}
\rho(x_{1},m_{1};x_{2},m_{2};x_{3},m_{3}) & = & \left \langle q^{m_{1}}\right
\rangle ^{j_{1}} \left \langle q^{m_{2}}\right
\rangle ^{j_{2}} \left \langle q^{m_{3}}\right
\rangle ^{j_{3}} \left \langle q^{m_{1}+m_{2}}\right
\rangle ^{j_{1,2}}\left \langle q^{m_{2}+m_{3}}\right
\rangle ^{j_{2,3}} \left \langle q^{m_{1}+m_{2}+m_{3}}\right
\rangle ^{j_{1,2,3}}.
\end{eqnarray} 
The first three factors constitute the multiplicative weights the three points do
not have in common, the next two factors constitute the multiplicative weights
which two
of the three points have in common and finally the last line counts the multiplicative
weights which all three points have in common.
The numbers of multiplicative weights are given analogous to (\ref{j12}) and (\ref{j1j2}), 
\begin{eqnarray}
&&j_{1}=\frac{V_{1} / V_{2}}{V_{\lambda}},\; j_{2}=\frac{V_{2}
/ (V_{1} \cup V_{3})}{V_{\lambda}},\; j_{3}=\frac{V_{3} / V_{2}}{V_{\lambda}} \\ & & \nonumber \\
&&j_{1,2}=\frac{(V_{1}\cap V_{2}) / V_{3}}{V_{\lambda}},\; j_{2,3} = \frac{(V_{2} \cap V_{3}) / V_{1}}{V_{\lambda}} \\ & & \nonumber \\
&&j_{1,2,3} = \frac{V_{1} \cap V_{2} \cap V_{3}}{V_{\lambda}}, 
\end{eqnarray} 
and are related to each other by
\begin{eqnarray}
\label{branch}
j_{1}+j_{1,2}+j_{1,2,3}= j_{2}+j_{1,2}+j_{2,3}+j_{1,2,3}= j_{3}+j_{2,3}+j_{1,2,3},
\end{eqnarray}
reflecting the total number of RMCP cascade steps. Higher order correlation densities are straightforward and result in the same
structure of common multiplicative weights of all combinations of neighbouring
points.

One more comment on the expression (\ref{three}). In discrete RMCP, say with
binary scale steps $\lambda=2$, three points first share a common cascade
history with $j_{1,2,3}$ steps. Then one point (say the third one) branches
off and from then on evolves independently from the other two; consequently,
$j_{2,3}=0$ could be zero and $j_{3}+j_{1,2,3}$ adds up to the total number of
cascade steps. Only $j_{1,2}$ could be different from zero, since the first
two points still share some more common cascade history; once those two
branches, $j_{1}=j_{2}$ independent steps are left until the RMCP evolution
reaches the dissipation scale $\eta$. However, this RMCP result,
i.e. $j_{1}+j_{1,2}+j_{1,2,3}=j_{2}+j_{1,2}+j_{1,2,3}=j_{3}+j_{1,2,3}$, is not
necessarily in conflict with the relation (\ref{branch}). The former reflects
the ultrametric view of the nested hierarchy of RMCP length scales and is not
observable for an experimentalist \cite{JOU99a,JOU00,JOU99b,CLE00,SCH,GRE98,EGG01}. An
unrestrictive sampling of discrete RMCP three-point correlations breaks the
underlying ultrametric structure, leading to a modified result which may be
closer to (\ref{branch}). This speculation should be tested with a discrete
RMCP simulation. Definitely even more interesting would be to test the
predicted generalised fusion rule structure (\ref{cd3}), (\ref{fusion3}) and
(\ref{three}) of three-point correlations with data. Already the two-point
correlations (\ref{twopo}), (\ref{cd2}), (\ref{fusion1}) and (\ref{mul1}) are
in excellent agreement with high Reynolds number data \cite{DZI01}, but
three-point correlations could represent an even more stringent model test. 
\subsection{Characteristic function}
In this subsection, we conclude our discussions on n-point statistics and
examine 
the most general space-time correlation-function $\langle
\epsilon(x_{1},t_{1})^{\xi_{1}} \cdot \ldots \cdot
\epsilon(x_{n},t_{n})^{\xi_{n}}\rangle$, containing all statistical
information. It is equal to the
generating function
\begin{eqnarray}
\label{gen}
Z[\xi(x,t)]=\left \langle\exp \left \{ \int_{-\infty}^{\infty}\!dx
\int_{-\infty}^{\infty}\!dt \;\xi(x,t) \; \ln \epsilon(x,t)\right \}   \right \rangle
\end{eqnarray}
for the logarithmic field, once 
$\xi(x,t)=\sum_{k=1}^{n}\xi_{k}\delta(t-t_{k})\delta(x-x_{k})$ is chosen. The
determination of $Z$ is easily performed by regarding (\ref{gen}) as a summation of
\emph{i.i.d.} stable-distributions $\gamma(x,t)$ integrated over their influence domain:
\begin{eqnarray}
\label{genfun}
Z[\xi(x,t)] & = & \left \langle \exp \left \{ \int_{-\infty}^{\infty}\!dx'
\int_{-\infty}^{\infty}\!dt' \gamma(x',t') V_{\xi}(x',t')\right \} \right
\rangle \nonumber \\ & & \nonumber  \\
 & = & \exp \left \{\frac{\sigma^{\alpha}}{\cos \left( \frac{\pi
\alpha}{2}\right)} \int_{-\infty}^{\infty}\!dx' \int_{-\infty}^{\infty}\!dt'
\left( V_{\xi}(x',t')-V^{\alpha}_{\xi}(x',t')\right)\right \}.
\end{eqnarray}
The first line is the result of inserting Eq.\ (\ref{def1}) into (\ref{gen});
also the abbreviation
\begin{equation}
\label{cf1}
V_{\xi}(x',t')=\int_{-\infty}^{\infty}\!dx \int_{-\infty}^{\infty}\!dt
f(x-x',t-t')\xi (x,t)
\end{equation}
has been introduced. For the second step of (\ref{genfun}), the Laplace transform
of stable distributions has been used.
The result (\ref{genfun}) is very convenient for numerical implementation: for
$n$-point correlation densities $\langle \epsilon(x_{1},t_{1})^{\xi_{1}}
\ldots \epsilon(x_{n},t_{n})^{\xi_{n}}\rangle$ the test function
$\xi(x,t)=\sum_{k=1}^{n} \xi_{k} \delta(x-x_{k}) \delta(t-t_{k})$ is a
weighted sum of $\delta$-functions, so that (\ref{cf1}) simply counts the
number of weighted spatio-temporal points inside the influence domain attached
to $(x',t')$. Note also, that the modified $n$-point correlation densities
$\langle (\ln \epsilon(x_{1},t_{1}))^{m_{1}} \cdots (\ln \epsilon(x_{n},t_{n}))^{m_{n}} \rangle$, which follow from (\ref{gen}) by taking
functional derivatives with respect to $\xi(x,t)$, only exist
for the lowest orders, once $\alpha \neq 2$ is chosen for the random field
$\gamma(x,t)$. 
\section{Coarse grained observables}
In this Section, we examine statistical properties of coarse grained
observables like integral moments and multiplier distributions. These are
representative observables for the analysis of the experimentally measured
surrogate energy dissipation field of fully developed turbulence.
\subsection{Integral moments}
The temporal integral moments of order $n$ are defined as the expectation
value of the coarse-grained temporal average of the
energy-dissipation field
\begin{eqnarray}
\label{mom}
M_{t}^{(n)}(x_{0},t_{0}) & = & \frac{1}{t^{n}}\left \langle\left(\int_{t_{0}-t/2}^{t_{0}+t/2} dt'
\epsilon(x_{0},t')\right)^{n} \right \rangle \sim \left(\frac{T}{t}\right)^{\tau_{t}(2)}.
\end{eqnarray}
Within the inertial range of fully developed turbulence these moments show multifractal scaling with exponent
$\tau_{t}(n)$ \cite{SRE97}. Spatial integral moments, together with spatial scaling
exponents $\tau_{x}(n)$ are defined as the coarse grained spatial average of
the energy-dissipation field. Invoking the 
Taylor-Frozen-Flow Hypothesis \cite{YAG71}, the spatial scaling exponents are believed to coincide with their temporal counterparts. For the time being, we
present analytical model results for the temporal coarse graining and show numerical results for
the spatial case.

Integral moments of order $n$ involve an integration over all
$n$-point correlations. This integration can be done by
integrating the second-order correlation function (\ref{cort}) with $n_{1}=n_{2}=1$:
\begin{eqnarray*}
M_{t}^{(2)}(x_{0},t_{0}) & = & c_{2}
\left(
\frac{t_{L}}{t}\right)^{2}+c_{1}\left(\frac{t_{L}}{t}\right)+c\left(
\frac{T-t_{\eta}}{t}\right)^{\tau_{2}}
\end{eqnarray*}
for $t_{L} \le t \le T-t_{\eta}$.
The exponent $\tau_{2}=a(2^{\alpha}-2)$ should be set equal to the
experimentally accepted value of $0.225$ \cite{SRE97}. For $t_{L}\ll t\ll
T-t_{\eta}$ the last term $\sim t^{-\tau_{2}}$ dominates the second-order integral
moment, so that $M_{t}^{(2)}(x_{0},t_{0}) \sim
t^{-\tau_{2}}$ for intermediate scales. The analytic expressions for the
coefficients $c$, $c_{1}$ and $c_{2}$ are:
\begin{eqnarray}
c & = & 2\exp \left \{\frac{\tau_{2}}{2a_{0}}V(T-t_{\eta})\right \}
\left ( \frac{1}{1-\tau_{2}}-\frac{1}{2-\tau_{2}}\right) \\ & &
\nonumber \\
c_{1} & = & 2\left \{ \frac{\exp \left\{\frac{\tau_{2}}{2a_{0}}V(t_{L})
\right\}}{1-2\tau_{2}} \left(2-2^{2\tau_{2}} \right)-\frac{\exp \left
\{\frac{\tau_{2}}{2a_{0}}V(T-t_{\eta}) \right
\}\left(T-t_{\eta}\right)^{\tau_{2}}t_{L}^{-\tau_{2}}}{1-\tau_{2}} \right\} \\
& & \nonumber
\\
c_{2} & = & 2 \left \{ \frac{\exp \left \{ \frac{\tau_{2}}{2a_{0}}V(t_{L})
\right \}}{1-2\tau_{2}}\left
(-\frac{2^{2\tau_{2}}}{2-2\tau_{2}}+\frac{4}{2-2\tau_{2}}-2 \right) +\frac{\exp
\left \{\frac{\tau_{2}}{2a_{0}}V(T-t_{\eta}) \right
\}\left(T-t_{\eta}\right)^{\tau_{2}}t_{L}^{-\tau_{2}}}{2-\tau_{2}}
\right \} 
\end{eqnarray}
with the $V(t)$-expressions given by (\ref{vol}).
The determination of higher order moments is straightforward, though cumbersome.

Fig. 5 shows the local
slope of $\log M^{(n)}$ as a function of the average domain $\log t$ and $\log
l$ for the temporal
(Fig. 5a) and the spatial integral moments
(Fig. 5b), respectively. Parameters have been set $\alpha=1.9$ and in
arbitrary units $t_{L}=5t_{\eta}=5$, $T=L=2^{8}$. For very small and very large scales there are weak
deviations from perfect scaling, which increase as the order $n$
increases. Comparing the 
temporal and spatial local slopes, there is clearly a larger scaling
regime for the spatial moments. This is a result of the small scale statistics where temporal and spatial
two-point correlations show a different behaviour. Once again, the message is
that it is better to use two-point correlators instead of integral moments to
extract multifractal scaling exponents.
\subsection{Multiplier-pdf}
Having examined moments of coarse-grained observables in the previous subsection, we now
investigate the probability density function $p(M)$ of ratios of these
observables, the so-called multipliers:
\begin{eqnarray}
\label{multiplier}
M_{x}(x,t,l_{x})& = & \frac{\int_{x-l_{x}/2}^{x} \epsilon(x',t) dx'}{\int_{x-l_{x}/2}^{x+l_{x}/2} 
\epsilon(x',t) dx'},\nonumber \\ & &   \\
M_{t}(x,t,l_{t})& = & \frac{\int_{t-l_{t}/2}^{t} \epsilon(x,t') dt'}{\int_{t-l_{t}/2}^{t+l_{t}/2} 
\epsilon(x,t') dt'}.\nonumber
\end{eqnarray}
Here we have only displayed the definition of a left-sided spatial and
temporal multiplier $M_{x}$ and $M_{t}$, respectively, where the daughter
domain of size $l/2$ is located at the left of the mother domain of size $l$. In the same
manner right-shifted or centered multipliers, also with other scale steps, can be
defined.
A series of experimental investigations \cite{SRE95,PED96} for large Reynolds
number turbulent flows resulted in  a $\beta$-distribution
with $\beta \sim 3.2$ for the left multiplier-distribution, scale-independent
in the upper part of the inertial range. The most important
result is reflected in the conditional multiplier distributions
$p(M(l)|M(2l))$ \cite{SRE95}, where correlations among multipliers with nested
coarse graining domains  at nearby scales have been observed:
compared to the unconditioned pdf, for the left/right-sided
multipliers one gets a more narrow pdf, if conditioned on a small parent multiplier  
and a more broader pdf, if conditioned on a large parent
multiplier. On first view,  
the conditioning effects violate the assumptions of RMCPs since those require
\emph{i.i.d.} random-weights $q(l)$. 
However, the contradiction is resolved once the observationally unavoidable
unrestrictive sampling and the assumption of a nonconservative cascade
generator is taken into account within the simple RMCPs \cite{JOU99a,JOU00}.
Since we pointed out that our dynamical approach can be
viewed as a generalisation of these successful models, we now ask, for the
outcome of 
these unconditioned and conditioned multiplier pdfs within the dynamical
cascade model. 

We also report the differences between
spatial and temporal multiplier distributions. The results presented here rely
on simulations. Model parameters have been set $\tau[1,1]=0.225$,
$\alpha=1.92$ and in arbitrary units
$T=L=2^{8}$,
$t_{\eta}=t_{L}=1$. The kernel-domain has been discretized into cells of size
$dt=0.1$, $dx=0.1$, each
filled with one random number $\gamma$. Stability of the
results with respect to cell size and the number of independent
samples of the energy-dissipation field has been checked. Fig. 6 shows (a) the
temporal and (b) the spatial multipliers (\ref{multiplier}) obtained from
$10^{5}$ field samples. For comparison the
$\beta$ distribution with $\beta=3.2$ is
depicted as the solid line. 

The unconditioned left-sided temporal multiplier-distribution is
well-fitted by the experimentally observed $\beta$-distribution, if the index
of stability $\alpha$ is set equal to $1.92$. For larger $\alpha$ all
distributions become slightly more narrow and for smaller $\alpha$ they become
slightly broader. Also the
correct scale-correlations are reproduced in the conditioned distributions.   These pdf's
are scale invariant for $l_{t}>2^{-4}T$, that is over four binary orders of
magnitude. For smaller scales all distributions become broader. 
The qualitative behaviour of temporal  multiplier distributions remains unchanged once
spatial multipliers are examined. But there is a clear quantitative
difference, all spatial multiplier distributions are a little broader compared to
their temporal counterpart. The differences arise again from the different small and
large scale behaviour of $n$-point correlations. Scale independence and conditioning effects are
nevertheless not affected. These model simulations demonstrate that the
proposed spatio-temporal cascade process is also able to reproduce the
experimental multiplier findings.
\section{Model generalisation to $\boldmath{3+1}$ dimensions}
So far the dynamical cascade process has been formulated in $1+1$
dimensions. Generalisations to $n+1$ dimensions are straightforward. Their
construction again relies on the scaling properties of spatial and temporal
two-point correlators. In the following we only state the results for the
special case $n=3$  where the boundary conditions yield the most simplest
analytical relations: 

Sticking to spherical symmetry, the kernel function $g(t)$
of $1+1$ dimensions is replaced by the time-dependent radius $r(t)$, so that
the scalar field
$\epsilon(\vec{r},t)$
\begin{eqnarray}
\epsilon(\vec{r},t) & = & \exp \left\{ \int_{t-T}^{t}dt'
\int_{B^{(3)}_{r(t'-t+T)}(\vec{r})} d\vec{r'} \gamma(\vec{r'},t)\right \}
\end{eqnarray}
is then composed of time-dependent $3$-dimensional
balls $B^{(3)}_{r(t)}(\vec{r})$.
Scaling relations for spatial and temporal two-point correlators with symmetric scaling exponents $\overline{\tau}=\tau$ 
immediately imply
\begin{eqnarray}
r(t) & = & \left \{ \begin{array}{ll} \left(
\frac{a_{0}}{t+\tilde{t}}-\frac{a_{0}}{T+\tilde{t}}\right)^{\frac{1}{3}} & \;\;\;\;\;\;T-t_{\eta}
\le t \le T\\ & \\
\left(\frac{a_{0}}{t}\right)^{\frac{1}{3}}    & \;\;\;\;\;\;t_{L} \le T-t_{\eta} \\& \\
\frac{a_{0}^{1/3}}{(t+t_{0})^{\xi}}& \;\;\;\;\;\;0 \le t \le t_{L}
\end{array} \right. .
\end{eqnarray}
The functional form for $t>T-t_{\eta}$ again is arbitrary and only affects a
constant prefactor in the temporal scaling. The functional form for $t_{L} \le
t \le T-t_{\eta}$ is fixed by the temporal scaling prerequisite $V(\Delta t) \sim \ln \Delta
t$. The new parameters $\xi$ and $t_{0}$ are fixed by two boundary conditions,
continuity at $t=t_{L}$ and a spatial scaling relation $V(\Delta x) \sim \ln
\Delta x$:
\begin{eqnarray}
t_{L}^{2\xi-2/3} & = & \frac{\left(
\frac{t_{0}}{t_{L}}+1\right)^{1-2\xi}-\left(
\frac{t_{0}}{t_{L}}\right)^{1-2\xi}}{3(1-2\xi)}, \nonumber \\ & & \\
t_{L} & = & \left(t_{0}+t_{L}\right)^{3 \xi}.  \nonumber
\end{eqnarray}
The choice $\xi=2$, for example results in $t_{L}\sim 0.33$ and
$t_{0}\sim 0.5$. Of course, other choices of $\xi$ are also possible.
\section{Conclusion}
The dynamical model presented here is constructed from a strict multifractal scaling of
two-point correlators and employs stable statistics of the random-field $\gamma$. The
latter allows an analytical treatment of $n$-point correlations. For
fully developed turbulence, the dynamical model is not only able to reproduce
the observed multifractal scaling of two-point correlators for the energy
dissipation field, but also the observed multiplier distributions, including
scale-correlations. The dynamic model also predicts specific generalised
fusion rules for $n$-point correlations; experimentally the sampling of at
least three-point correlations should be feasible and could represent an even
more stringent test of the proposed model. Another interesting observable, to
be discussed in future, are $n$-point cumulants of logarithmic field
amplitudes \cite{EGG01}. In this context, it might be necessary to formulate
the proposed continuous dynamical model in a discrete way, introducing
smallest spatio-temporal cells of size of the order of the dissipation scale
and thus allowing to include deviations from log-stable statistics.

Already in the last Section, we have sketched the formulation of the dynamical
model in $3+1$ dimensions. The extension to arbitrary $n+1$ dimensions is
straightforward. Other generalisations in $n+1$ dimensions would be to include
spatial anisotropy and boundary layer effects; work in this direction is
already in progress \cite{SCH01}. As possible applications of the proposed
dynamical cascade model, we see simulations for the spatio-temporal evolution
of multifractal fields like rain and cloud fields in geophysics. In a
numerically improved way, it might even serve as an efficient model generator
for the turbulent small scale motion entering large-eddy simulations and other
numerical approaches.

\acknowledgements
The authors acknowledge fruitful discussions with Hans C. Eggers. This work
has been supported in parts by DAAD.

\newpage
\begin{figure}
\begin{centering}
\epsfig{file=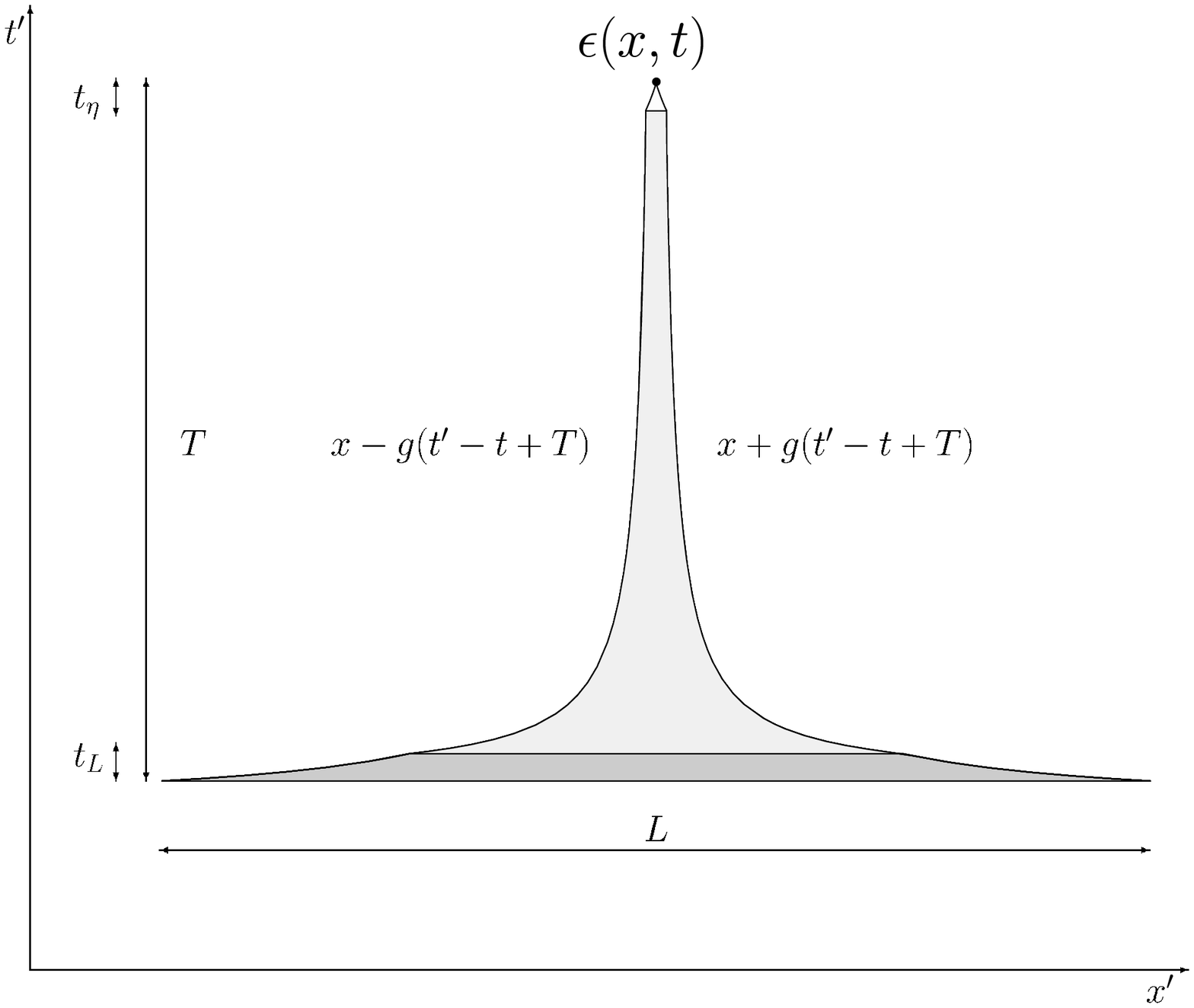,width=15cm}
\caption{Illustration of the causal index function $f$ and the kernel function
$g(t)$ used to construct the
positive-valued multifractal field $\epsilon(x,t)$.
} 
\end{centering}
\end{figure}
\newpage
\begin{figure}
\begin{centering}
\epsfig{file=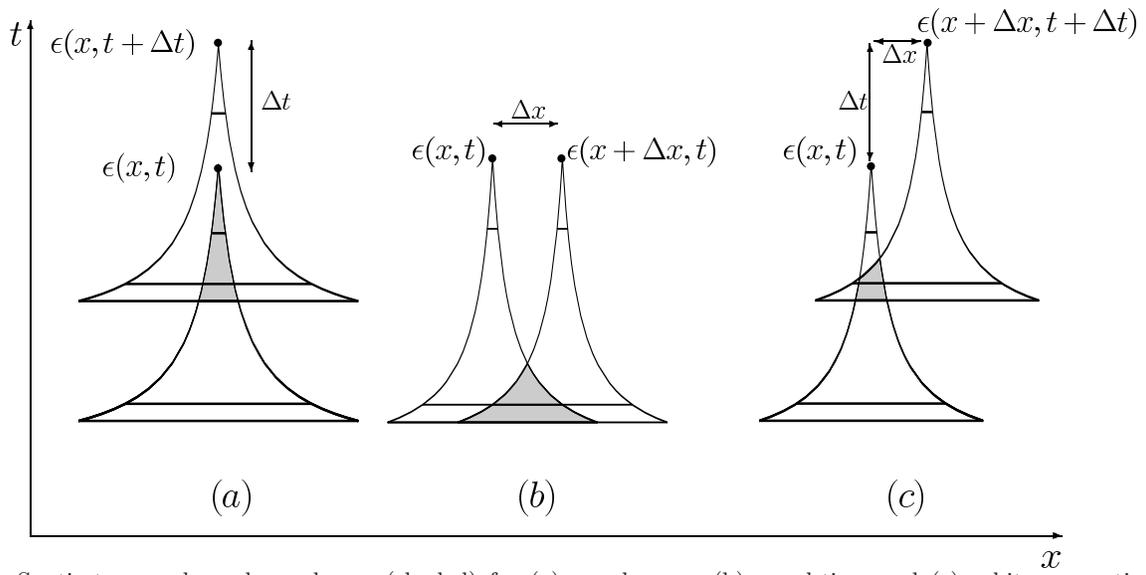,width=15cm}
\caption{Spatio-temporal overlap volumes (shaded) for (a) equal-space, (b) equal-time and (c) arbitrary spatio-temporal
distance.} 
\end{centering}
\end{figure}
\newpage
\begin{figure}
\begin{centering}
\epsfig{file=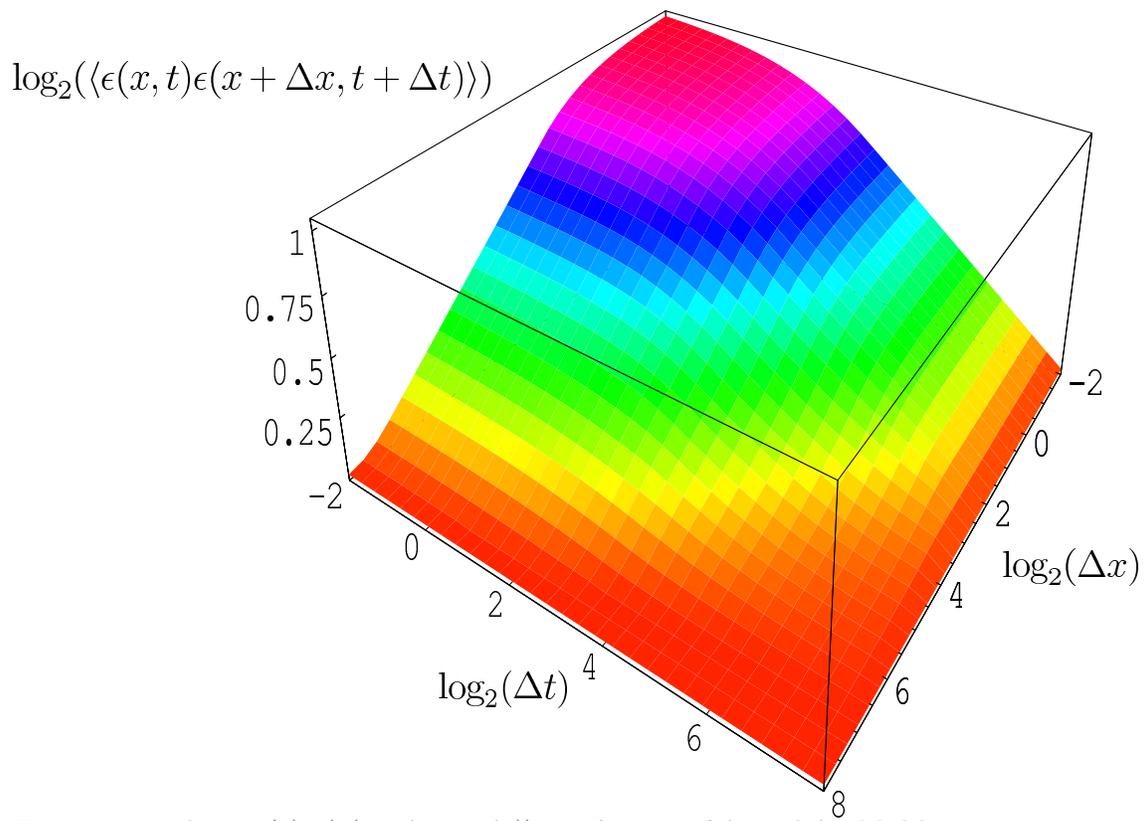,width=15cm}
\caption{Two-point correlations
$\left \langle \epsilon(x,t)\epsilon(x+\Delta x,t+\Delta t) \right \rangle$ as
a function of $\Delta x$ and $\Delta t$. Model parameters are quoted in the text.}
\end{centering}
\end{figure}
\newpage
\begin{figure}
\begin{centering}
\epsfig{file=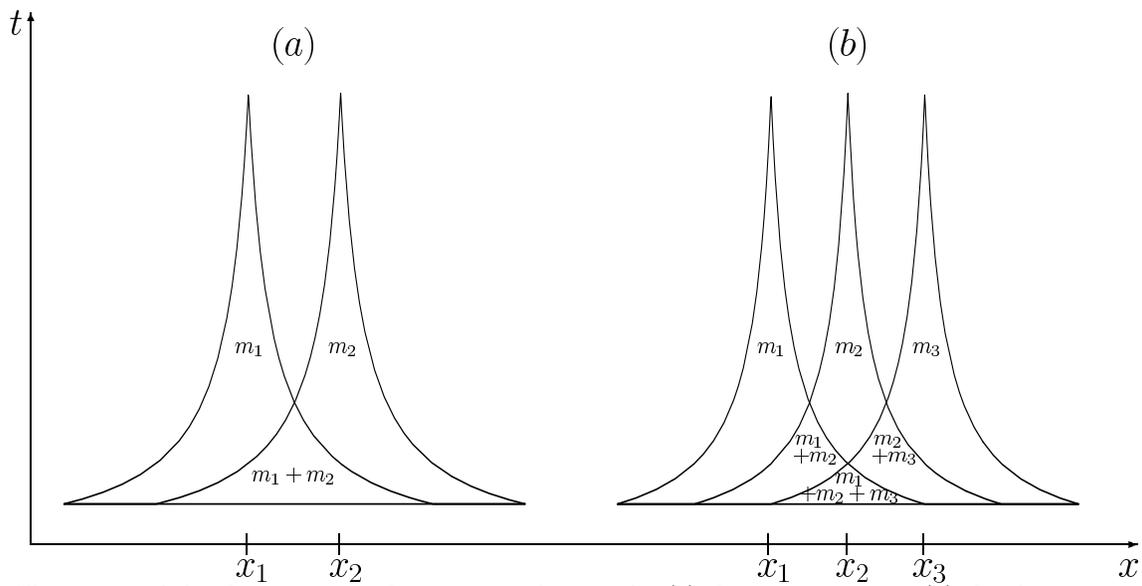,width=15cm}
\caption{Illustration of the three resp.\ six disjunct contributions for (a)
the two-point resp.\ (b) the three-point equal-time correlation density.}
\end{centering}
\end{figure}
\newpage
\begin{figure}
\begin{centering}
\epsfig{file=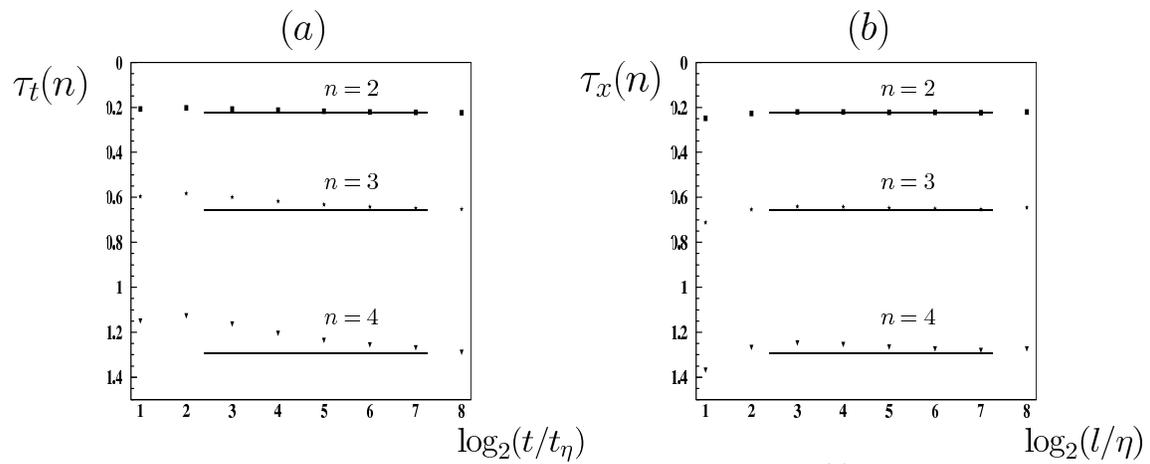,width=15cm}
\caption{Local slopes of (a) the temporal and (b) the spatial integral moments
$M^{(n)}$ of order $n=2,3,4$ as a function of the coarse-graining scale $t$
and $l$, respectively. Model parameters are quoted in the text. For comparison the two-point scaling
exponents $\tau_{n}=0.225(n^{\alpha}-n)/(2^{\alpha}-2)$ are indicated by
straight lines.} 
\end{centering}
\end{figure}
\newpage
\begin{figure}
\begin{centering}
\epsfig{file=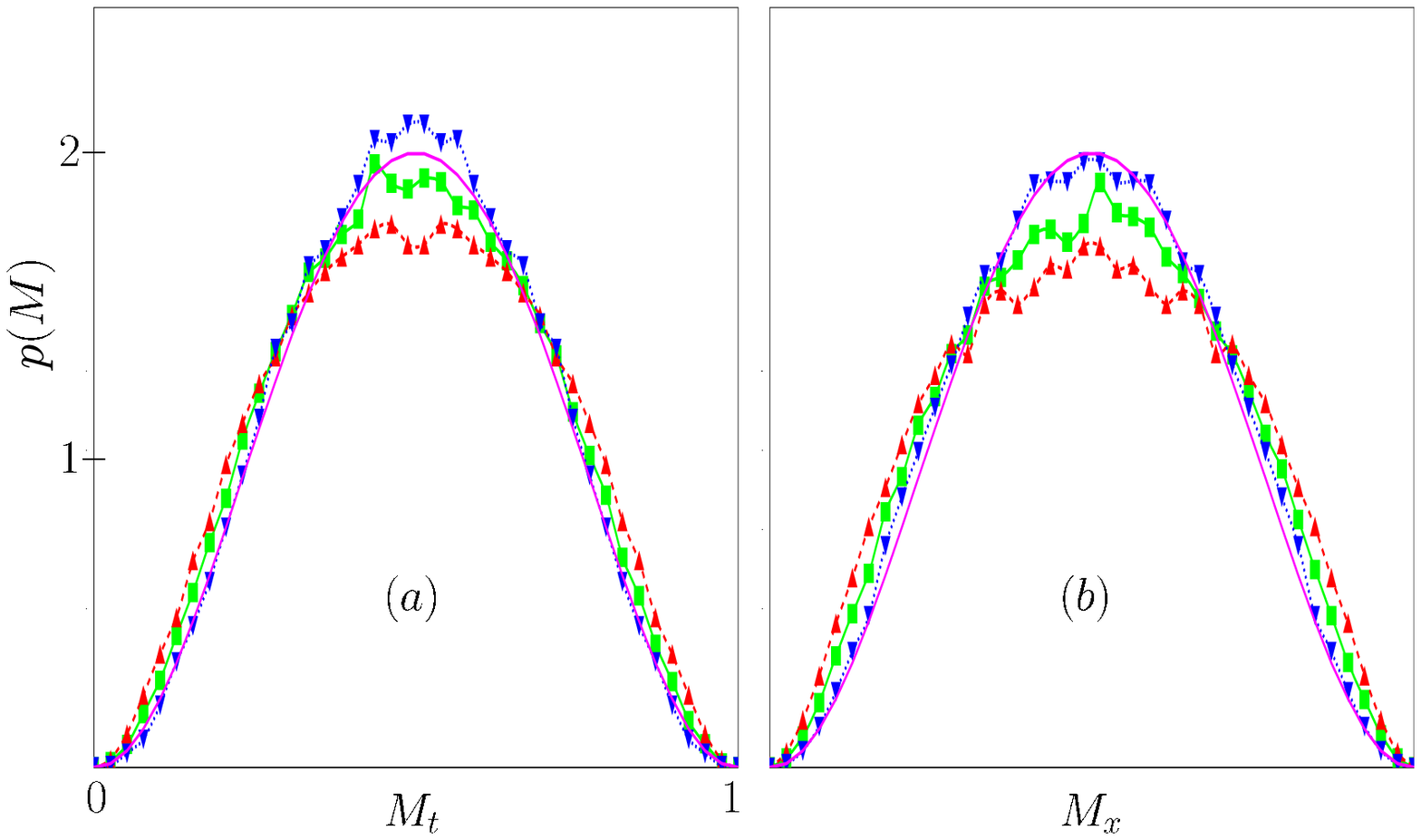,width=15cm}
\caption{$(a)$ temporal and $(b)$ spatial multiplier-distributions $p(M(l))$
($\blacksquare$),   $p(M(l)|M(2l)<0.5)$ ($\blacktriangledown$),   $p(M(l)|M(2l)>0.5)$ ($\blacktriangle$) for $t=T/8$ and $l=L/8$. Model parameters as quoted in the text.}
\end{centering}
\end{figure}

\newpage
\begin{thebibliography}{99}
\bibitem{YAG71}
         A.S.\ Monin and A.M.\ Yaglom, 
         {\em Statistical Fluid Mechanics}, Vols.\ 1 and 2, 
         (MIT Press, Cambridge, 1971).
\bibitem{KOL41}
	A.N.\ Kolmogorov, 
	Dokl.\ Akad.\ Nauk.\ SSSR {\bf 32}, 16 (1941).
\bibitem{FRI95}
        U.\ Frisch, {\em Turbulence}, 
	(Cambridge University Press, Cambridge, 1995). 
\bibitem{SRE97}
         K.R.\ Sreenivasan and R.A.\ Antonia,
         Ann.\ Rev.\ Fluid Mech.\ {\bf 29}, 435 (1997).
\bibitem{BOHR98}
	T.\ Bohr, M.H.\ Jensen, G.\ Paladin and A.\ Vulpiani, {\em Dynamical
	systems approach to turbulence}, 
	(Cambridge University Press, Cambridge, 1998).
\bibitem{RICH22}
        L.F.\ Richardson, {\em Weather prediction by numerical process}, 
	(Cambridge University Press, Cambridge, 1922).
\bibitem{MEN91}
	C.\ Meneveau and K.R.\ Sreeenivasan,
	J.\ Fluid\ Mech.\ {\bf 224}, 429 (1991).
\bibitem{SRE95}
         K.R.\ Sreenivasan and G.\ Stolovitzky,
         J.\ Stat.\ Phys.\ {\bf 78}, 311 (1995).
\bibitem{MOL95}
         J.\ Molenaar, J.\ Herweijer and W.\ van de Water, 
         Phys.\ Rev.\ E {\bf 52}, 496 (1995).
\bibitem{PED96}
         G.\ Pedrizzetti, E.A.\ Novikov and A.A.\ Praskovsky,
         Phys.\ Rev.\ E\ {\bf 53}, 475 (1996).
\bibitem{NAE}
         A.\ Naert, R.\ Friedrich, and J.\ Peinke,
         Phys. \ Rev.\ E {\bf 56}, 6719 (1997).
\bibitem{JOU99a}
         B.\ Jouault, P.\ Lipa, and M.\ Greiner,
         Phys. \ Rev.\ E {\bf 59}, 2451 (1999).
\bibitem{JOU00}
         B.\ Jouault, M.\ Greiner, and P.\ Lipa,
         Physica D {\bf 136}, 125 (2000).
\bibitem{JOU99b}
         B.\ Jouault, J.\ Schmiegel, and M.\ Greiner,
         chao-dyn/9909033.
\bibitem{CLE00}
         J.\ Cleve and M.\ Greiner,
         Phys.\ Lett.\ A {\bf 273}, 104 (2000).
\bibitem{let}
        J.\ Schmiegel, H.\ C.\ Eggers and M.\ Greiner, preprint 
	cond-mat/0106347.
\bibitem{SCHM}
        F.\ Schmitt and D.\ Marsan, preprint 
	cond-mat/0102346.
\bibitem{TAQ94}
	G.\ Samarodnitzky and M.\ Taqqu, {\em Stable non-Gaussian random
        processes} , (Chapman \& Hall, New York, 1994).
\bibitem{GRE96}
         M.\ Greiner, J.\ Giesemann, P.\ Lipa and P.\ Carruthers, 
         Z.\ Phys.\ C {\bf 69}, 305 (1996).
\bibitem{WOL00}
         M.\ Wolf and M.\ Greiner,
         Phys.\ Lett.\ A {\bf 266}, 276 (2000).
\bibitem{SCH}
	J.\ Schmiegel, T.\ Dziekan, J.\ Cleve, B.\ Jouault and M.\ Greiner,
	preprint {\em Scaling functions in a random multiplicative
	energy-cascade model of turbulence}.
\bibitem{PRO96}
	V.\ L'vov and I.\ Procaccia,
	Phys.\ Rev.\ Lett.\ {\bf 76}, 2898 (1996).
\bibitem{GRE98}
        M.\ Greiner, J.\ Giesemann und P.\ Lipa,
        Phys.Rev.\ E56 4263 (1997).
\bibitem{EGG01}
        H.C.\ Eggers, T.\ Dziekan and M.\ Greiner,
        Phys.\ Lett.\ A {\bf 281}, 249 (2001).
\bibitem{DZI01}
        T.\ Dziekan,"Zwei-Punkt-Kumulanten in synthetischer Turbulenz",
        diploma thesis (TU Dresden, 2001);
        T.\ Dziekan, J.\ Cleve and K.R.\ Sreenivasan, private communication.
\bibitem{SCH01}
        J.\ Schmiegel and M.\ Greiner, in preparation.  
\end{thebibliography}
\end{document}